\documentclass[aps, pra, reprint, floatfix, superscriptaddress, showkeys, nofootinbib, fleqn]{revtex4-2}
\pdfoutput=1
\usepackage{amsmath, amsfonts, amssymb, amsthm, bm, bbm}
\usepackage[colorlinks=true, linkcolor=blue, citecolor=magenta, urlcolor=blue]{hyperref}
\usepackage{lmodern}
\usepackage[T1]{fontenc}
\usepackage[utf8]{inputenc}
\usepackage[american]{babel}
\usepackage[activate={true,nocompatibility},final,tracking=true,kerning=true,spacing=true,factor=1100]{microtype}
\usepackage{orcidlink}
\usepackage{titlesec}
\usepackage{nccmath}
\usepackage{tikz-cd}
\usepackage{multirow, booktabs}
\usepackage{tikz}

\usepackage[margin=1.4cm]{geometry}

\theoremstyle{remark}
\newtheorem{rmk}{Remark}

\newtheorem*{rmk*}{Remark}
\newtheorem*{ex*}{Example}

\newcommand{\HH}{\mathcal{H}}
\newcommand{\RR}{\mathcal{R}}
\newcommand{\EE}{\mathcal{E}}

\newcommand{\GG}{\mathcal{G}}
\newcommand{\VV}{\mathcal{V}}

\newcommand{\UU}{\mathrm{U}}

\newcommand{\tr}{\operatorname{tr}}
\newcommand{\dt}{\mathrm{d}t}
\newcommand{\bra}[1]{\langle #1|}
\newcommand{\ket}[1]{|#1\rangle}
\newcommand{\braket}[2]{\langle #1 | #2 \rangle}
\newcommand{\ketbra}[2]{|#1 \rangle\langle #2|}
\newcommand{\llangle}{\langle\!\langle}
\newcommand{\rrangle}{\rangle\!\rangle}

\newcommand{\linspan}{\operatorname{span}}

\newcommand{\length}{\operatorname{Length}}

\newcommand{\boldsigma}{\boldsymbol{\sigma}}
\newcommand{\boldOmega}{\boldsymbol{\Omega}}

\newcommand{\bolda}{\boldsymbol{a}}
\newcommand{\boldr}{\boldsymbol{r}}

\begin{document}

\title{Isoholonomic inequality and tight implementations of holonomic quantum gates}
\author{Ole S{\"o}nnerborn\,\orcidlink{0000-0002-1726-4892}\,}
\email{ole.sonnerborn@kau.se}
\affiliation{Department of Mathematics and Computer Science, Karlstad University, 651 88 Karlstad, Sweden}
\affiliation{Department of Physics, Stockholm University, 106 91 Stockholm, Sweden}
\date{May 17, 2025}

\begin{abstract}
In holonomic quantum computation, quantum logic gates are realized by cyclic parallel transport of the computational space. The resulting quantum gate corresponds to the holonomy associated with the closed path traced by the computational space. The isoholonomic inequality for gates establishes a fundamental lower bound on the path length of such cyclic transports, which depends only on the spectrum of the holonomy, that is, the eigenvalues of the implemented quantum gate. The isoholonomic inequality also gives rise to an estimate of the minimum time required to execute a holonomic quantum gate, underscoring the central role of the inequality in quantum computation. In this paper, we show that when the codimension of the computational space is sufficiently large, any quantum gate can be implemented using a parallel transporting Hamiltonian in a way that saturates the isoholonomic inequality and the corresponding time estimate. We call such implementations tight. The treatment presented here is constructive and lays the foundation for the development of efficient and optimal implementation strategies in holonomic quantum computation.
\end{abstract}


\maketitle

\titleformat{\section}[block]{\bfseries\large}{\Roman{section}}{0.7em}{}
\titlespacing{\section}{0em}{1.2em}{1em}
\titleformat{\subsection}[block]{\bfseries\normalsize}{\Roman{section}.\Alph{subsection}}{0.7em}{}
\titlespacing{\subsection}{0em}{1.2em}{0.5em}
\titleformat{\subsubsection}[block]{\itshape\normalsize}{\Roman{section}.\Alph{subsection}.\arabic{subsubsection}}{0.4em}{}
\titlespacing{\subsubsection}{0em}{0.5em}{0.3em}

\section{Introduction}
\label{sec: Introduction}
\noindent
Non-adiabatic parallel transport has been proposed as an efficient technique for implementing quantum logic gates that combines the robustness of adiabatic holonomic quantum computation with enhanced resilience to noise \cite{ZaRa1999, SjToAnHeJoSi2012, XuZhToSjKw2012, SjMoCa2016, ZhKyFiKwSjTo2023}. 

A key result in holonomic quantum computation is the isoholonomic inequality, which establishes a fundamental lower bound on the path length necessary for a cyclic transformation of the computational space to realize a holonomic gate. This inequality also provides an estimate of the minimum time required to execute a holonomic quantum gate \cite{So2024a}. In this paper, we first review the principles of nonadiabatic holonomic quantum computation and formulate the isoholonomic inequality and its associated execution time estimate. We then show that if the codimension of the computational space is at least as large as its dimension, any quantum gate can be tightly implemented using a parallel transporting Hamiltonian. Here, ``tightly'' signifies that the isoholonomic inequality and the execution time estimate are saturated. The proof is constructive and explicit, making it a suitable starting point for the design of tight implementation schemes for holonomic quantum gates.

\section{Parallel transport and holonomic gates}
\label{sec: Parallel transport and holonomic gates}
\noindent 
In nonadiabatic holonomic computation, gates are implemented by cyclic parallel transport of the computational space within an ambient Hilbert space. To accurately describe this procedure we need to introduce the Stiefel-Grassmann principal bundle.

\subsection{The Stiefel-Grassmann bundle}
\label{sec: The Stiefel-Grassmann bundle}
\noindent
Let $\HH$ be a Hilbert space and $n$ be a positive integer less than the dimension of $\HH$. The Grassmann manifold $\GG(n;\HH)$ is the manifold of all $n$-dimensional subspaces of $\HH$. We provide this manifold with the smooth and Riemannian structures induced by its canonical identification with the space of orthogonal projection operators of rank $n$ on $\HH$, equipped with the Hilbert-Schmidt Riemannian metric.

Assume the computational space $\RR$ is a subspace of $\HH$ of dimension $n$, and thus is an element of $\GG(n;\HH)$. A cyclic transformation of $\RR$ corresponds to a curve in $\GG(n;\HH)$ starting and ending at $\RR$.\footnote{In this paper we use the word \emph{curve} to refer to a piecewise smooth one-parameter family of objects such as vectors, operators, or subspaces. We call the parameter time, denote it by $t$, and assume that it varies between $0$ and some unspecified finite value $\tau$.} In this paper we focus exclusively on cyclic transformations generated by Hamiltonians, that is, closed curves in $\GG(n;\HH)$ of the form $\RR_t = U_t(\RR)$, where $U_t$ is the time evolution operator of a given Hamiltonian. 

An $n$-frame in $\HH$ is a sequence of $n$ orthonormal vectors in $\HH$, and the Stiefel manifold $\VV(n;\HH)$ is the manifold of all such frames. The linear span of an $n$-frame is an element of $\GG(n;\HH)$, and the surjective map 
\begin{equation}
    \VV(n;\HH)\ni V\mapsto \linspan{V}\in\GG(n;\HH)
\end{equation}
is the projection of the Stiefel-Grassmann principal bundle.

The symmetry group of the Stiefel-Grassmann bundle is the group $\UU(n)$ of $n\times n$ unitary matrices, acting on $\VV(n; \mathcal{H})$ from the right. To describe this action more explicitly we represent the elements of $\VV(n; \mathcal{H})$ as formal row matrices, 
\begin{equation}
    V=\big(\ket{v_1}\, \ket{v_2}\dots\ket{v_n}\big).
\end{equation}
The action of a unitary matrix $U$ on $V$ is then given by the matrix multiplication $V \mapsto VU$. Since this multiplication replaces each component of $V$ by a linear combination of the components of $V$ and thereby preserves its span, the action of the symmetry group preserves the fibers of the Stiefel-Grassmann bundle.

\begin{rmk}
    In this paper, all frames are assumed to be of length $n$ unless explicitly stated otherwise. Therefore, from now on, we will simply write \emph{frame} and omit the reference to the number of components.
\end{rmk}   

\subsection{Parallel transport}
\label{sec: Parallel transport}
\noindent
Parallel transport ensures that the gates implemented are the result of purely geometric effects. To rigorously define the concept of parallel transport, we must first define what is meant by a curve of frames being horizontal.  

We say that a tangent vector $\dot V$ at a frame $V$ is horizontal if the velocity vectors that make up $\dot V$ are perpendicular to the span of $V$. In short, $\dot V$ is horizontal if $V^\dagger \dot V=0$. Furthermore, we say that a curve of frames $V_t$ is horizontal if its velocity vector is always horizontal,
\begin{equation}
    V_t^\dagger \dot V_t=0.
    \label{eq: horizontal curve of frames}
\end{equation}
A fundamental result from the theory of principal fiber bundles ensures that for every curve $\RR_t$ in $\GG(n; \HH)$ and every frame $V$ spanning $\RR_0$, there exists a unique horizontal curve of frames $V_t$ with $\linspan{V_t}=\RR_t$ and $V_0=V$. We call $V_t$ the horizontal lift of $\RR_t$ starting at $V$.

Suppose $\RR_t = U_t(\RR)$, where $U_t$ is the evolution operator of a Hamiltonian $H_t$. For any frame $V$ spanning $\RR$, we can define a lift $V_t$ of $\RR_t$ as $V_t = U_t V$, where $U_t$ acts component\-wise on $V$. This lift can be horizontal or not. If it is horizontal, that is, if condition \eqref{eq: horizontal curve of frames} is satisfied at all times, we say that $H_t$ is parallel transporting. The property of being parallel transporting does not depend on the choice of frame $V$ but is intrinsic to $H_t$ itself. By writing $\ket{v_{k;t}}$ for the $k$th component of $V_t$, the parallel transport condition can be written as
\begin{equation}
    \bra{v_{k;t}} H_t \ket{v_{l;t}} = 0 \quad (k,l=1,2,\dots,n).
    \label{eq: parallel transport condition}
\end{equation}

\subsection{Holonomic gates}
\label{sec: Holonomic gates}
\noindent
Suppose $\RR_t$ is a closed curve at $\RR$, and $V_t$ is an arbitrary horizontal lift of $\RR_t$. Then the two frames $V_0$ and $V_\tau$ span $\RR$. As a result, the unitary operator
\begin{equation}
    \Gamma[\RR_t]=V_\tau V_0^\dagger = \sum_{k=1}^n \ketbra{v_{k;\tau}}{v_{k;0}}
\end{equation}
preserves $\mathcal{R}$. This operator is independent of the choice of the horizontal lift and is called the holonomy of $\RR_t$. We define a holonomic gate to be any unitary transformation of the computational space $\RR$ that is implemented as the holonomy of a parallel transport of $\RR$ around a loop. If the parallel transport is unitary, $\RR_t=U_t(\RR)$, then the restriction of the final time unitary $U_\tau$ to $\RR$ is equal to the holonomy of $\RR_t$ and is thus a holonomic gate. This is how holonomic gates are implemented in practice.

\begin{rmk}
    If the parallel transport condition is satisfied, $\RR$ is transported in such a way that no time-local rotation occurs within $\RR$. Nevertheless, any set of cotransported vectors in $\RR$ may end up rotated relative to its initial configuration. The rotation is described by the holonomy.      
\end{rmk}

\subsection{Projective holonomic gates}
\label{sec: Projective holonomic gates}
\noindent
Two gates differing only by a phase factor transform states in an identical way and are therefore computationally interchangeable. Consequently, one can more generally define a gate on $\RR$ to be a projective unitary operator on $\RR$ or, equivalently, a holomorphic isometry of the projective space of $\RR$. Reference \cite{So2024b} developed a gauge theory for nonadiabatic holonomic quantum computation with projective gates. The projective parallel transport condition is given by 
\begin{equation}
    \bra{v_{k;t}} H_t \ket{v_{l;t}} = \epsilon_t\delta_{kl} \quad (k,l=1,2,\dots,n),
\end{equation}
where $\epsilon_t$ is any real-valued function. We recover the conventional parallel transport condition by requiring that $\epsilon_t=0$.

\section{The isoholonomic inequality}
\label{sec: The isoholonomic inequality}
\noindent 
The isoholonomic bound of a gate $\Gamma$ on an $n$-dimensional subspace of $\HH$ is the quantity
\begin{equation}
    L(\Gamma) = \sqrt{ \sum_{j=1}^n \theta_j ( 2\pi - \theta_j ) },
\end{equation}
where $\theta_1, \theta_2, \dots, \theta_n$ are the phases of the eigenvalues of $\Gamma$ in the interval $[0, 2\pi)$. The isoholonomic inequality for gates states that the length of a closed curve $\RR_t$ in $\GG(n;\HH)$ is lower bounded by the isoholonomic bound of its holonomy: 
\begin{equation}
    \length[\RR_t] \geq L\big(\Gamma[\RR_t]\big).
    \label{eq: the isoholonomic inequality}
\end{equation}
The length of $\RR_t$ is defined as 
\begin{equation}
    \length[\RR_t] = \int_0^\tau \dt\, \sqrt{ \frac{1}{2} \tr \left( \dot P_t^2 \right)}, 
\end{equation}
where $P_t$ is the orthogonal projection onto $\RR_t$. The isoholonomic inequality for gates was derived in Ref.\ \cite{So2024a}.

From the isoholonomic inequality one can derive a quantum speed limit for the execution time of holonomic gates: Suppose $\RR_t$ is generated by the Hamiltonian $H_t$. Then the length of $\RR_t$ can be expressed as 
\begin{equation}
    \length[\RR_t] = \tau\llangle \sqrt{I(H_t;\RR_t)} \, \rrangle.
    \label{eq: the length}
\end{equation}
The factor in double angle brackets is the average of the square root of the skewness measure
\begin{equation} 
    I(H_t;\RR_t) = - \frac{1}{2} \tr\big( [H_t, P_t]^2 \big)
\end{equation}
over the evolution time interval. The skewness measure quantifies both (the square of) the instantaneous speed with which $\RR_t$ traverses Hilbert space and the extent to which $\RR_t$ is not invariant under $H_t$; see Ref.\ \cite{LuSu2020}. It follows from Eqs.\ \eqref{eq: the isoholonomic inequality} and \eqref{eq: the length} that if $H_t$ is configured to parallel transport the computational subspace along the closed curve $\RR_t$, thereby executing a holonomic gate $\Gamma$, then the execution time is not less than the isoholonomic bound of $\Gamma$ divided by the mean value of the square root of the skewness measure:
\begin{equation} 
    \tau \geq \frac{ L(\Gamma) }{ \llangle \sqrt{ I(H_t;\RR_t) }\,\rrangle }. 
    \label{eq: the conventional quantum speed limit}
\end{equation}

\subsection{The projective isoholonomic inequality}
\label{sec: The projective isoholonomic inequality}
\noindent
There is also a projective version of the isoholonomic inequality which states that the length of any closed curve $\RR_t$ in $\GG(n;\HH)$ is bounded from below by the isoholonomic bound of its projective holonomy:
\begin{equation}
    \length[\RR_t]\geq L\big(\bar{\Gamma}[\RR_t]\big).
\end{equation}
The isoholonomic bound of a projective gate $\bar{\Gamma}$ is defined as the minimum of the isoholonomic bounds of the gates representing $\bar{\Gamma}$:
\begin{equation}
    L(\bar{\Gamma})=\min\{L(\Gamma):\Gamma\in\bar{\Gamma}\}.
\end{equation}
It was shown in Ref.\ \cite{So2024b} that if $\Gamma$ is an arbitrary representative of $\bar{\Gamma}$, $\theta_1,\theta_2,\dots,\theta_n$ are the phases in $[0,2\pi)$ of the eigenvalues of $\Gamma$, and $\theta_0=0$, then
\begin{equation}
    L(\bar{\Gamma})=\min_{0\leq k\leq n}\bigg{\{} \sum_{j=1}^n |\theta_j-\theta_k|\big(2\pi-|\theta_j-\theta_k|\big) \bigg{\}}.
    \label{eq: the projective isoholonomic bound}
\end{equation}
The isoholonomic inequality for projective gates gives rise to a quantum speed limit for the execution time of projective gates similar to that in Eq.\ \eqref{eq: the conventional quantum speed limit}.

\subsection{Review of the derivation of the isoholonomic inequality}
\label{sec: Review of the derivation of the isoholonomic inequality}
\noindent
The isoholonomic inequality for gates \eqref{eq: the isoholonomic inequality} extends and generalizes the isoholonomic inequality for states derived in Ref.\ \cite{HoSo2023a} (see also the Supplemental Material in Ref.\ \cite{So2024a}). The latter inequality asserts that if $\rho_t$ represents a closed curve of pure states, then its Fubini–Study length is bounded from below as follows: 
\begin{equation} 
    \mathrm{Length}[\rho_t] \geq \sqrt{\theta(2\pi - \theta)}. 
    \label{eq: isoholonomic inequality for states} 
\end{equation} 
The quantity $\theta$ is the Aharonov-Anandan geometric phase of $\rho_t$ in the interval $[0,2\pi)$.\footnote{The holonomy of a closed curve of pure states is an element in $\UU(1)$, and the Aharonov-Anandan geometric phase is defined as the argument of this holonomy modulo $2\pi$; see \cite{AhAn1987}. The phase $\theta$ is the unique representative of this argument in the interval $[0,2\pi)$.} 

Let $V_t$, with components $\ket{v_{k;t}}$, be a horizontal lift of $\RR_t$ initiated at an eigenbasis for the holonomy of $\RR_t$:
\begin{equation} 
    \Gamma[\RR_t]\ket{v_{k;0}} = e^{i\theta_k}\ket{v_{k;0}}. 
\end{equation} 
Each curve of pure states $\rho_{k;t}=\ketbra{v_{k;t}}{v_{k;t}}$ is then closed with the geometric phase $\theta_k$. Furthermore it holds that 
\begin{equation} 
    \mathrm{Length}[\RR_t]^2 \geq \sum_{k=1}^n \mathrm{Length}[\rho_{k;t}]^2. 
\end{equation} 
See Ref.\ \cite{So2024a} for details. The isoholonomic inequality for gates is obtained by applying the isoholonomic inequality for states to each of the terms on the right.

\section{Tight implementations}
\label{sec: Time optimal implementations}
\noindent
In this section we show that the isoholonomic inequality is tight when the codimension of the computational subspace is at least as large as its dimension. More precisely, we show that under these conditions it is possible (at least in principle) to implement any quantum gate on the computational space using a parallel transporting Hamiltonian that transports the computational space along a cyclic path whose length saturates the isoholonomic inequality and hence the quantum speed limit \eqref{eq: the conventional quantum speed limit}. We call such an implementation a tight implementation.

\subsection{Tight parallel generation of a geometric phase}
\label{sec: Tight generation of a geometric phase}
\noindent
The proof of the isoholonomic inequality suggests that we first focus on tight parallel transports of a state that generates a given geometric phase. Let $\ket{v}$ be a normalized vector in $\HH$ and $\EE$ be any two-dimensional subspace of $\HH$ containing $\ket{v}$. Inspired by Ref.\ \cite{HoSo2023b}, we show that there exists a Hamiltonian $H_t$ which acts trivially on the orthogonal complement of $\EE$ and parallel transports $\ket{v}$ along a curve $\ket{v_t}$ with the following properties:
\begin{itemize}
    \item $\ket{v_t}$ remains within $\EE$, and for an arbitrarily chosen time $\tau>0$ we have that $\ket{v_\tau}=e^{i\theta}\ket{v}$, where $\theta$ is any predetermined phase in the interval $[0,2\pi)$.
    \item $\rho_t=\ketbra{v_t}{v_t}$ is periodic with period $\tau$ and has the constant Fubini-Study speed $(2\pi\theta-\theta^2)^{1/2}/\tau$.
\end{itemize}
Since $H_t$ is parallel transporting, that is, $\bra{v_t}H_t\ket{v_t}=0$, each loop of $\rho_t$ acquires the geometric phase $\theta$. Moreover, the speed of $\rho_t$ is calibrated so that each loop saturates the isoholonomic inequality for states. The ability to confine the dynamics to a given two-dimensional subspace will be crucial in the following sections.

To construct a Hamiltonian that satisfies the above criteria let $\ket{\epsilon_0},\ket{\epsilon_1}$ be an orthonormal basis for $\EE$ such that
\begin{equation}
    2\pi|\braket{\epsilon_1}{v}|^2=\theta,
    \label{eq: crucial overlap}
\end{equation}
and define the triple of Pauli-like operators $\boldsigma$ as 
follows:
\begin{subequations}
\begin{align}
    \sigma_1 &= \ketbra{\epsilon_0}{\epsilon_1} + \ketbra{\epsilon_1}{\epsilon_0}, \label{eq: Pauli-x}\\
    \sigma_2 &= i( \ketbra{\epsilon_0}{\epsilon_1} - \ketbra{\epsilon_1}{\epsilon_0} ), \label{eq: Pauli-y}\\
    \sigma_3 &= \ketbra{\epsilon_1}{\epsilon_1} - \ketbra{\epsilon_0}{\epsilon_0}. \label{eq: Pauli-z}
\end{align}
\end{subequations}
The Bloch vector of $\rho=\ketbra{v}{v}$ is the unit-length $3$-vector $\boldr$ with components $r_k=\bra{v}\sigma_k\ket{v}$. Define $A$ and $H$ as
\begin{subequations}
\begin{alignat}{3}
    A &= \bolda\cdot\boldsigma,\quad && \bolda &&= (0,0,\pi/\tau), \label{eq: A Rabi vector} \\ 
    H &= \boldOmega\cdot\boldsigma,\quad && \boldOmega &&= \bolda-(\bolda\cdot\boldsymbol{r})\boldsymbol{r}. \label{eq: H Rabi vector}
\end{alignat}
\end{subequations}
The claim is that the Hamiltonian 
\begin{equation}
    H_t = e^{-itA} H e^{itA}
    \label{eq: tight qubit Hamiltonian}
\end{equation}
meets the required criteria.

Let $U_t$ be the time evolution operator associated with $H_t$, and set $\ket{v_t}=U_t\ket{v}$. The time evolution operator can be decomposed as $U_t=e^{-itA} e^{-it(H-A)}$, and the condition \eqref{eq: H Rabi vector} ensures that $\rho$ commutes with $H-A$:
\begin{equation} 
    [H - A, \rho] = 2i \big( (\boldOmega - \bolda) \times \boldr \big) \cdot \boldsigma = 0.
\end{equation} 
As a result, $\rho$ evolves as if $A$ were the Hamiltonian: 
\begin{equation} 
    \rho_t = U_t \rho U_t^\dagger = e^{-itA} \rho e^{itA}. 
    \label{eq: effective Hamiltonian}  
\end{equation} 
Condition \eqref{eq: H Rabi vector} also ensures that $H_t$ is parallel transporting:
\begin{equation} 
    \bra{v_t} H_t \ket{v_t} = \tr(H_t \rho_t) = \tr(H \rho) = \boldOmega \cdot \boldr = 0. 
\end{equation}
From Eqs.\ \eqref{eq: A Rabi vector} and \eqref{eq: effective Hamiltonian} it follows that $\rho_t$ is periodic with the period $\pi/|\bolda|=\tau$. Furthermore, each loop acquires the Aharonov-Anandan geometric phase $\theta$:
\begin{equation}
    \pi \big(1 + \bra{v} \sigma_3 \ket{v} \big) 
    = 2\pi |\braket{\epsilon_1}{v}|^2 = \theta. 
\end{equation} 
Here we have applied a well-known formula for the geometric phase of a qubit whose dynamics is governed by a time-independent Hamiltonian (see, e.g., \cite{Sj2015}), which is effectively the case here according to Eq.\ \eqref{eq: effective Hamiltonian}.

By construction, $H_t$ acts trivially on the orthogonal complement of $\EE$, ensuring that $\ket{v_t}$ remains in $\EE$. Since $H_t$ is parallel transporting, $\ket{v_t}$ is Aharonov-Anandan horizontal. We also have that $\ket{v_\tau}=e^{i\theta}\ket{v}$ since $\rho_t$ completes a loop with the geometric phase $\theta$ in the time $\tau$.

It remains to calculate the Fubini-Study speed of $\rho_t$. The square of the speed is
\begin{equation}
    \frac{1}{2}\tr\big(\dot{\rho}_t^2\big)
    = -\frac{1}{2}\tr\big([A,\rho]^2\big)
    = |\bolda\times\boldr|^2.
\end{equation}
Furthermore,
\begin{equation}
\begin{split}
    \tau^2|\bolda\times\boldr|^2 
    &= \tau^2|\bolda|^2\big(1-\bra{v}\sigma_3\ket{v}^2\big) \\
    &= \pi^2\big(1-(2|\braket{v}{\epsilon_1}|^2-1)^2\big) \\
    &= 2\pi|\braket{v}{\epsilon_1}|^2\big(2\pi-2\pi|\braket{v}{\epsilon_1}|^2\big) \\
    &= \theta(2\pi-\theta).
\end{split}
\end{equation}
We conclude that $\rho_t$ has a constant speed given by
\begin{equation}
    \sqrt{\frac{1}{2}\tr\big(\dot{\rho}_t^2\big)}
    =|\bolda\times\boldr|
    =\frac{1}{\tau}\sqrt{\theta(2\pi-\theta)}.
\end{equation}

\begin{figure}[t]
    \centering
    \includegraphics[width=.30\textwidth]{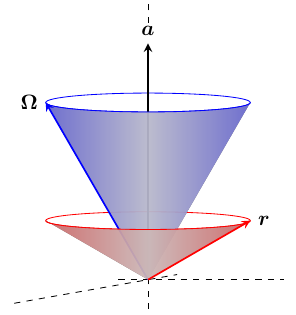}
    \caption{The Bloch vector $\boldr$ of the state and the Rabi vector $\boldOmega$ of the Hamiltonian are perpendicular at $t=0$. These vectors rotate with time around the axis spanned by the vector $\bolda$, representing the operator $A$. The rotation is illustrated by cones. The red cone is traced out by the Bloch vector of the evolving state, and the blue cone is traced out by the Rabi vector of the time-varying Hamiltonian. The Bloch and Rabi vectors rotate with the same angular speed and therefore remain perpendicular.}
    \label{fig: Bloch figure}
\end{figure}

Figure \ref{fig: Bloch figure} illustrates the perpendicular configuration of the Bloch vector $\boldr$ and the Rabi vector $\boldOmega$ representing the state and the Hamiltonian at the time $t=0$. As time passes, the Bloch and Rabi vectors of the state and the Hamiltonian rotate with the same angular speed, $2|\bolda|$, around the axis spanned by $\bolda$, and thus remain perpendicular. 

\begin{rmk}
    If $\theta=\pi$, then $H_t=A$. This is the only nonzero geometric phase of a qubit system that can be implemented by a time-independent parallel transporting Hamiltonian. For all other nonzero phases, the Hamiltonian must be time-dependent.
    \label{rmk: phase pi}
\end{rmk}

\subsection{Tight implementation of a holonomic gate that has only one eigenvalue with a non-zero phase}
\label{sec: Tight implementation of a holonomic gate that has only one eigenvalue with a non-zero phase}
\noindent
Section \ref{sec: A complete set of qubit gates} provides schemes for the tight implementation of the $\pi/8$, Hadamard, and CNOT gates. Common to these is the property that they can be tightly implemented with a Hamiltonian that acts nontrivially on only one eigenvector of the gate to be implemented. As an application of the previous section, we outline an approach for this case. The general case will be treated in detail in the next section. 

Suppose the gate to be implemented, $\Gamma$, has only one eigenvalue whose phase $\theta$ is not zero. Let $\ket{v}$ be a corresponding normalized eigenvector. To implement $\Gamma$ tightly, the Hamiltonian only has to act nontrivially on the linear span of this vector and a vector $\ket{w}$ in the orthogonal complement of the computational space. We follow the strategy of the previous section and set 
\begin{subequations}
\begin{align}
    \ket{\epsilon_0} &= \sqrt{1-\frac{\theta}{2\pi}}\ket{v} - \sqrt{\frac{\theta}{2\pi}}\ket{w}, \\
    \ket{\epsilon_1} &= \sqrt{\frac{\theta}{2\pi}}\ket{v} + \sqrt{1-\frac{\theta}{2\pi}}\ket{w}.
\end{align}
\end{subequations}
The overlap condition \eqref{eq: crucial overlap} is then satisfied. We define Pauli-like operators as in Eqs.\ \eqref{eq: Pauli-x}--\eqref{eq: Pauli-z}, and we define $A$ and $H$ as in Eqs.\ \eqref{eq: A Rabi vector} and \eqref{eq: H Rabi vector}, with the Bloch vector being that of $\ketbra{v}{v}$. The Hamiltonian is then given by
\begin{equation}
    H_t = e^{-itA}H e^{itA}.
\end{equation}
This Hamiltonian parallel transports $\ket{v}$ to $e^{i\theta}\ket{v}$ within the span of $\ket{v}$ and $\ket{w}$, while keeping the orthogonal complement of $\ket{v}$ in the computational space fixed. The Hamiltonian thus parallel transports the computational space in a closed curve with holonomy $\Gamma$. The length of this curve is the same as the Fubini-Study length of the evolution curve of $\ketbra{v}{v}$, which is $(2\pi\theta-\theta^2)^{1/2}$. The length thus matches the isoholonomic bound of $\Gamma$, which means that the implementation is tight.

\subsection{Tight implementation of a holonomic gate}
\label{sec: Tight implementation of a holonomic gate}
\noindent
Assume the codimension of the computational space is at least equal to its dimension. One can then apply a tight transformation to each eigenstate of the target gate without simultaneously coupling it to other eigenstates. In the following, we show that this selective approach allows a tight holonomic implementation of any desired gate.

Consider a target gate $\Gamma$ acting on the computational space $\RR$, which we aim to implement tightly using a parallel transporting Hamiltonian $H_t$. Let $\theta_1,\theta_2,\dots,\theta_n$ be the phases of the eigenvalues of $\Gamma$, chosen within the interval $[0,2\pi)$, and let $V$ be any frame of corresponding eigenvectors for $\Gamma$,
\begin{equation}
    V=\big(\ket{v_1}\, \ket{v_2}\cdots \ket{v_n}\big),\quad \Gamma\ket{v_k}=e^{i\theta_k}\ket{v_k}.
\end{equation}
Furthermore, let $\ket{w_1}, \ket{w_2},\dots,\ket{w_n}$ be an orthonormal set of vectors in the orthogonal complement of $\RR$. As established in the previous section, for each eigenvector $\ket{v_k}$ there exists a Hamiltonian $H_{k;t}$ that evolves $\ket{v_k}$ along a trajectory $\ket{v_{k;t}}$ within the two-dimensional subspace $\EE_k$ spanned by $\ket{v_k}$ and $\ket{w_k}$. The Hamiltonian is configured so that the projection $\rho_{k;t}=\ketbra{v_{k;t}}{v_{k;t}}$ traces a closed path over a fixed time $\tau$, thereby accumulating the geometric phase $\theta_k$, and the Hamiltonian leaves the orthogonal complement of $\EE_k$ unaffected. We define the full Hamiltonian as 
\begin{equation}
    H_t =   \sum_{k=1}^n H_{k;t}.
    \label{eq: full Hamiltonian}
\end{equation}

Let $U_t$ be the time evolution operator of $H_t$. The curve of frames $V_t=U_tV$ has the $\ket{v_{k;t}}$s as components and is a horizontal lift of $\RR_t=\linspan V_t$. This follows from $H_t$ parallel transporting the eigenstates of $\Gamma$ within mutually orthogonal subspaces of $\HH$. Also, at time $t=\tau$, the subspace $\RR_t$ returns to $\RR$ because $\ket{v_{k;\tau}}=e^{i\theta_k}\ket{v_{k}}$, implying that $V_\tau$ spans the same space as $V$. The holonomy of $\RR_t$ is given by
\begin{equation}
    U_\tau\big|_{\RR}=V_\tau V^\dagger = \sum_{k=1}^n e^{i\theta_k} \ketbra{v_k}{v_k} = \Gamma.
\end{equation}

To determine the length of $\RR_t$, we start by calculating the skewness of $H_t$ relative to $\RR_t$:
\begin{equation}
\begin{split}
    I(H_t;\RR_t) 
    &= -\frac{1}{2}\tr\big([H_t,V_tV_t^\dagger]^2\big) \\
    &= \sum_{k=1}^m -\frac{1}{2}\tr\big([H_{k;t},\ketbra{v_{k;t}}{v_{k;t}}]^2\big) \\
    &= \sum_{k=1}^m \frac{1}{2}\tr\big(\dot{\rho}_{k;t}^2\big) \\
    &= \frac{1}{\tau^2}\sum_{k=1}^m \theta_k(2\pi-\theta_k).
\end{split}
\end{equation}
In the second line we have used the fact that $H_t$ preserves the mutually orthogonal subspaces $\mathcal{E}_k$. We find that the length of $\RR_t$ corresponds to the isoholonomic bound of $\Gamma$:
\begin{equation}
\begin{split}
    \length[\RR_t]^2 
    &= \tau^2 I(H_t;\RR_t) \\
    &= \sum_{k=1}^n \theta_k(2\pi-\theta_k) \\
    &= L(\Gamma)^2.
\end{split}
\end{equation}
This shows that the implementation of $\Gamma$ is tight.

\section{A complete set of qubit gates}
\label{sec: A complete set of qubit gates}
\noindent
A complete set of qubit gates allows the approximation of any quantum gate with arbitrary precision on a computational space associated with a register of qubits. The single-qubit $\pi/8$ and Hadamard gates, together with the two-qubit CNOT gate, form a complete set \cite{Bo2000}. Here we provide tight holonomic implementation schemes for these gates.

\subsection{The $\pi/8$ gate}
\noindent
Let $\ket{0},\ket{1}$ be the computational basis of a qubit. The $\pi/8$ gate $\mathrm{T}$, which adds $\pi/4$ to the relative phase between these basis vectors, is the gate represented by the matrix
\begin{equation}
    \mathrm{T}=\begin{pmatrix} 
    1 & 0 \\ 0 & e^{i\pi/4} 
    \end{pmatrix}.
\end{equation}
This gate has eigenvalues $1$ and $e^{i\pi/4}$, with phases $0$ and $\pi/4$. Consequently, the isoholonomic bound of the $\pi/8$ gate is
\begin{equation}
    L(\mathrm{T})=\sqrt{\frac{\pi}{4}\Big(2\pi-\frac{\pi}{4}\Big)} = \frac{\pi\sqrt{7}}{4}.
\end{equation}

To implement $\mathrm{T}$ holonomically and tightly in the time $\tau$, we need to find a Hamiltonian that tightly parallel transports $\ket{1}$ to $e^{i\pi/4}\ket{1}$ in the time $\tau$ while keeping $\ket{0}$ fixed. Guided by the approach described in Sec.\ \ref{sec: Tight implementation of a holonomic gate that has only one eigenvalue with a non-zero phase}, we fix a unit vector $\ket{2}$ in the orthogonal complement of the computational space and introduce the orthonormal vectors
\begin{subequations}
\begin{align}
    \ket{\epsilon_0} &= \sqrt{\frac{7}{8}}\,\ket{1} - \sqrt{\frac{1}{8}}\,\ket{2}, \\
    \ket{\epsilon_1} &= \sqrt{\frac{1}{8}}\,\ket{1} + \sqrt{\frac{7}{8}}\,\ket{2}.
\end{align}
\end{subequations}
The choice of coefficients is such that Eq.\ \eqref{eq: crucial overlap} is satisfied. We define $A$ and $H$ as in Eqs.\ \eqref{eq: A Rabi vector} and \eqref{eq: H Rabi vector}. Then $H_t=e^{-itA}He^{itA}$ tightly implements $\mathrm{T}$ in the time $\tau$. The operators $A$ and $H$ are represented by the matrices
\begin{subequations}
\begin{align}
    A &= \frac{\pi}{4\tau}
    \begin{pmatrix} 
    -3 & \sqrt{7} \\ 
    \sqrt{7} & 3
    \end{pmatrix},
    \\
    H &= \frac{\pi}{4\tau}
    \begin{pmatrix} 
    0 & \sqrt{7} \\ 
    \sqrt{7} & 0
    \end{pmatrix},
\end{align}
\end{subequations}
relative to the frame $\ket{1},\ket{2}$.

\begin{rmk}
Sometimes the $\pi/8$ gate is defined as 
\begin{equation}
    \mathrm{T}'= \begin{pmatrix} 
    e^{-i\pi/8} & 0 \\ 0 & e^{i\pi/8} 
    \end{pmatrix}.
\end{equation}
The phases in $[0,2\pi)$ of the eigenvalues of this matrix are $\pi/8$ and $15\pi/8$. Thus, the isoholonomic bound of $\mathrm{T}'$ is
\begin{equation}
    L(\mathrm{T}')
    = \frac{\pi}{4}\sqrt{\frac{15}{2}},
\end{equation}
which is greater than the isoholonomic bound of $\mathrm{T}$. Both $\mathrm{T}$ and $\mathrm{T}'$ represent the same projective gate, but one must transport the computational space along a longer curve to implement $\mathrm{T}'$ holonomically compared to $\mathrm{T}$. It follows from Eq.\ \eqref{eq: the projective isoholonomic bound} that the isoholonomic bound of the projective gate represented by $\mathrm{T}$ and $\mathrm{T}'$ is the same as that of $\mathrm{T}$.
\end{rmk}

\subsection{The Hadamard gate}
\noindent 
The Hadamard gate is the qubit gate represented by
\begin{equation}
    \mathrm{H}= \frac{1}{\sqrt{2}}
    \begin{pmatrix} 
    1 & 1 \\ 1 & -1
    \end{pmatrix}
\end{equation}
relative to a computational basis $\ket{0},\ket{1}$. The Hadamard gate has eigenvalues $1$ and $-1$, with phases $0$ and $\pi$, implying that the isoholonomic bound of the Hadamard gate is
\begin{equation}
    L(\mathrm{H}) = \sqrt{\pi(2\pi-\pi)} = \pi.
\end{equation}

An eigenvector of $\mathrm{H}$ with eigenvalue $-1$ is
\begin{align}
    \ket{v}&=\sin\Big(\frac{\pi}{8}\Big)\ket{0}-\cos\Big(\frac{\pi}{8}\Big)\ket{1}.
\end{align}
We fix a unit vector $\ket{2}$ in the orthogonal complement of the computational space and define
\begin{subequations}
\begin{align}
    \ket{\epsilon_0} &= \frac{1}{\sqrt{2}}\big(\ket{v} - \ket{2}\big), \\
    \ket{\epsilon_1} &= \frac{1}{\sqrt{2}}\big(\ket{v} + \ket{2}\big).
\end{align}
\end{subequations}
Then the overlap condition \eqref{eq: crucial overlap} is satisfied. Since the geometric phase to be implemented is $\pi$, we set $H_t=A$ as per Remark \ref{rmk: phase pi}. With respect to the frame $\ket{0},\ket{1},\ket{2}$,
\begin{equation}
    A = \frac{\pi}{\tau}
    \begin{pmatrix} 
    0 & 0 & {\hspace{9pt}}\sin(\pi/8) \\
    0 & 0 & -\cos(\pi/8) \\ 
    \sin(\pi/8) & -\cos(\pi/8) & 0
    \end{pmatrix}
\end{equation}

\begin{rmk}
    The projective gate represented by $\mathrm{H}$ has the same isoholonomic bound as $\mathrm{H}$.
\end{rmk}

\subsection{The CNOT gate}
\noindent 
Suppose the computational space is spanned by the two-qubit computational basis $\ket{00},\ket{01},\ket{10},\ket{11}$. The CNOT gate is the entangling two-qubit gate
\begin{equation}
    \mathrm{CNOT}=\begin{pmatrix}
        1 & 0 & 0 & 0 \\
        0 & 1 & 0 & 0 \\
        0 & 0 & 0 & 1 \\
        0 & 0 & 1 & 0
    \end{pmatrix}.
\end{equation}
The CNOT gate has eigenvalues $1$ and $-1$, the former with multiplicity $3$. Thus, the isoholonomic bound of CNOT is
\begin{equation}
    L(\mathrm{CNOT})=\sqrt{\pi(2\pi-\pi)} = \pi.
\end{equation}

An eigenvector with eigenvalue $-1$ is
\begin{equation}
    \ket{v}= \frac{1}{\sqrt{2}}\big(\ket{10}-\ket{11}\big).
\end{equation}
In complete analogy to the case of the Hadamard gate, CNOT can be tightly implemented with a time-independent Hamiltonian $H_t=A$, which in this case is given by the matrix
\begin{equation}
    A = \frac{\pi}{2\tau}
    \begin{pmatrix} 
    0 & 1 & 0 \\
    1 & 0 & -\sqrt{2} \\ 
    0 & -\sqrt{2} & 0
    \end{pmatrix}
\end{equation}
relative to $\ket{10},\ket{11},\ket{2}$. As before, $\ket{2}$ is an arbitrary normalized vector in the orthogonal complement of the computational space.

\section{Summary}
\label{sec: Summary}
\noindent
The isoholonomic inequality establishes a lower bound on the path length required to parallel transport the computational space for implementing a quantum gate holonomically. This inequality is a cornerstone of holonomic quantum computation, as it not only imposes fundamental constraints but also provides an estimate of the minimum time needed to execute a holonomic gate. As such, it is crucial for designing fast and efficient schemes for holonomic quantum computation.

In this paper, we have demonstrated that the isoholonomic inequality and its associated execution time estimate are both sharp and saturable. Specifically, we have proven that any quantum gate can be implemented using a Hamiltonian that parallel transports the computational space along a loop that exactly saturates these bounds. Our presentation, being both constructive and explicit, paves the way for the development of optimal and efficient strategies for implementing holonomic quantum gates.

\end{document}